\documentclass[german,english]{bvm2020}

\def\articlenumber{0000}

\date{}
\title{Fooling the Crowd with Deep Learning-based Methods}

\author{Christian~Marzahl$^1{}^,{}^2$,
Marc~Aubreville$^1$, 
Christof~A.~Bertram$^3$, 
Stefan~Gerlach$^2$, 
Jennifer~Maier$^1$,
J{\"o}rn~Voigt$^2$, 
Jenny Hill$^4$, 
Robert~Klopfleisch$^3$, 
Andreas~Maier$^1$}

\authorrunning{Marzahl et al.}

\institute{
$^1$Pattern Recognition Lab, Department of Computer Science, Friedrich-Alexander-Universit{\"a}t 
$^2$Research \& Development Projects, EUROIMMUN Medizinische Labordiagnostika AG \\
$^3$Institute of Veterinary Pathology, Freie Universit{\"a}t Berlin, Germany \\
$^4$VetPath Laboratory Services, Ascot,Western Australia
}

\email{c.marzahl@euroimmun.de}

\begin{document}

%
\selectlanguage{english}

\maketitle

\begin{abstract}
Modern, state-of-the-art deep learning approaches yield human like performance in numerous object detection and classification tasks. The foundation for their success is the availability of training datasets of substantially high quantity, which are expensive to create, especially in the field of medical imaging. Recently, crowdsourcing has been applied to create large datasets for a broad range of disciplines. This study aims to explore the challenges and opportunities of crowd-algorithm collaboration for the object detection task of grading cytology whole slide images. We compared the classical crowdsourcing performance of twenty participants with their results from crowd-algorithm collaboration. All participants performed both modes in random order on the same twenty images.
Additionally, we introduced artificial systematic flaws into the precomputed annotations to estimate a bias towards accepting precomputed annotations. We gathered 9524 annotations on 800 images from twenty participants organised into four groups in concordance to their level of expertise with cytology. The crowd-algorithm mode improved on average the participants' classification accuracy by 7\%, the mean average precision by 8\% and the inter-observer Fleiss' kappa score by 20\%, and reduced the time spent by 31\%. However, two thirds of the artificially modified false labels were not recognised as such by the contributors. This study shows that crowd-algorithm collaboration is a promising new approach to generate large datasets when it is ensured that a carefully designed setup eliminates potential biases.
\end{abstract}

\section{Introduction}
In recent years, the field of computer vision has experienced tremendous improvements in object detection and classification, largely due to to the availability of high quality, high quantity labelled image databases and fostered by deep learning technologies. In the medical image domain, the availability of such data sets is still limited for many tasks, as expert-labelled data sets are expensive to create.
To explore the potential of reducing the human annotation effort while maintaining expert-level quality, we reviewed a method called crowd-algorithm collaboration where humans manually correct labels precomputed by an automatic system. In contrast to classical crowdsourcing and its numerous successful applications ~\cite{0000-03}, this new method of crowdsourcing has been rarely applied to medical datasets, for example by Maier-Hein et al.~\cite{0000-01} and Ganz et al.~\cite{0000-02}. 
In the present study, we aimed to investigate several research questions regarding crowd-algorithm collaboration for labelling cytology datasets: First, can crowdsourcing be applied to grade cells and is there a minimum skill level required? Second, what are the advantages and disadvantages of crowd-algorithm collaboration? Third, can the crowd be fooled by artificially modified annotations and if so, by what type of modifications? Finally, what would be a promising design for using crowdsourcing for whole slide image annotation? 
To achieve our aims, we carefully designed, performed and evaluated a set of experiments with crowdsourcing and crowd-algorithm collaboration on a pulmonary haemorrhage cytology dataset. 
This dataset was selected because it is a realistic example for the medical field due to its properties of having a high inter- and intra-observer variability and only a few examples explaining the grading process in the reference paper~\cite{0000-05}. A trained deep learning model to create algorithmic annotations developed by Marzahl et al.~\cite{0000-04} is publicly available making the dataset suitable for the presented crowd-algorithm collaboration study.

\section{Material and methods}

Our research group built a dataset of 57 cytological slides of equine bronchoalveolar lavage fluid. The slides were prepared by cytocentrifugation and then stained to highlight the cellular iron content with Prussian Blue (n=28) or Turnbull's Blue (n=29), which result in an identical colour pattern. The glass slides were digitalised using a linear scanner (Aperio ScanScope CS2, Leica Biosystems, Germany) at a magnification of 400$\times$ with a resolution of $ 0.25 ~\frac{\mu m}{px}$. Finally, 17 slides were completely annotated and scored by a veterinary pathologist, according to M. Y. Doucet and L. Viel~\cite{0000-06} into five grades from zero to four. This annotated part of the dataset containing 78,047 labeled cells ($\mu$ = 4,591, $\sigma$  = 3,389).

\subsection{Patch selection}

To evaluate the human inter- and intra-observer variability jointly with the influence of precomputed annotations, we extracted twenty 256$\times$256 pixels patches from the unannotated slides. According to Irshad  et al.~\cite{0000-07}, the crowdsourcing performance degrades significantly with larger patch sizes. In consequence, we used the smallest reasonable patch size, which can contain 15 of the largest cells. Twenty patches with at least 15 cells each contain around 300 hemosiderophages as recommended for grading by Golde et al. ~\cite{0000-07}. The patches were chosen such that each patch covered all on that whole slide image (WSI) available grades, that only one patch was extracted per WSI, and that the two staining types were equally represented over all patches.

\subsection{Label generation}
For the crowd-algorithm mode, the labels were generated by the RetinaNet implementation provided from \cite{0000-04} on the same twenty images. To investigate the effect of augmented images, the predictions were modified on some images. On five images, we removed the augmented annotation for one cell. On five other images, we increased all grades by one step. Finally, five images contained standard object detection-caused artefacts like multiple annotations for one cell or false positive hemosiderophages.

\subsection{Labelling platform}

\begin{figure}[hbt!]
\includegraphics[height=5.7cm]{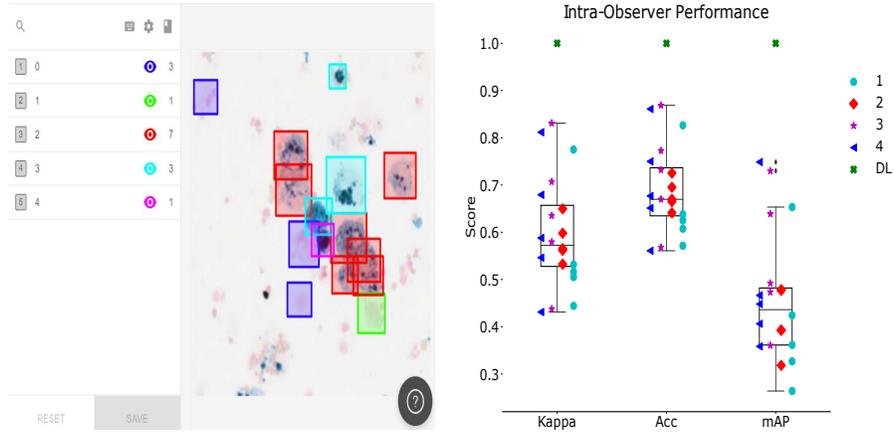}
\caption{From left to right: The screenshot shows the Labelbox\cite{0000-08} user interface with the precomputed annotations; On the right, the intra-observer performance for Cohens Kappa, classification accuracy (Acc) and mean average precision (mAP) for the groups of participants and the deep learning approach (DL). }
\label{0000-fig1}
\end{figure}

In order to estimate the effect of human qualification and experience, we divided our twenty contributors into four groups according to their qualification and experience with bronchoalveolar lavage (BAL) cytology. 
\begin{enumerate}
\item No experience in cytology (e.g computer scientist or chemists)
\item Beginner skills in cytology (e.g. biologist in training)
\item Professionals in the field of cytology (e.g. trained biologist)
\item Veterinary pathologist or clinician with a high degree of experience in BAL cytology.
\end{enumerate}
All contributors have provided written consent to participate in this study.  
We employed the Labelbox~\cite{0000-08} platform to host our experiments. Labelbox is a crowdsourcing platform which focuses on combining human annotations and machine learning methods to create high-quality datasets. Fig. ~\ref{0000-fig1} left visualises the LabelBox annotation interface. Anonymity was ensured by checking that no private information is saved in the files' meta-data and that no personal information can be extracted.

\subsection{Label experiment design}

We designed our experiments with the aim of estimating the effect of computer-aided annotation methods on crowdsourcing. For that purpose, two modes were created in Labelbox: Crowd-algorithm mode, where the contributor is asked to enhance the predictions made by a deep learning system, and annotation mode, where all annotations were performed by the participants without algorithmic support.

\section{Results}

In total, the twenty contributors made 9524 annotations on 800 images which took around 20 hours. Three veterinary pathologists defined the ground truth by majority vote. Additionally, we compared the contributors' performance with an algorithmic baseline set by a customised RetinaNet model \cite{0000-04}.

\begin{figure}[hbt!]
\includegraphics[width=1\textwidth]{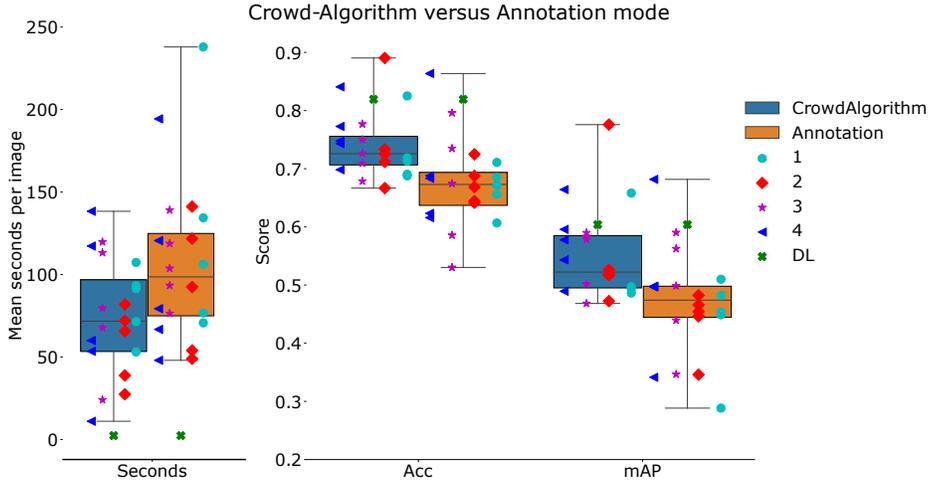}
\caption{The crowd-algorithm versus annotation plot compares the increased Acc and mAP jointly with the decreased interaction time between groups for the crowd-algorithm mode. DL represents the deep learning-based scores and the numbers from one to four denote the groups of participats.}
\label{0000-fig2}
\end{figure}

The crowd-algorithm mode led to better results on average than the annotation mode with an accuracy ranging from 67-89\% ($\mu$=74, $\sigma$=6) compared to 53-86\% ($\mu$=67, $\sigma$=7), while the deep learning-based approach alone reached an accuracy of 86\% (Fig. ~\ref{0000-fig2}). The inter-observer Fleiss' kappa score was 0.51 for the annotation mode and 0.71 for the crowd-algorithm mode. In crowd-algorithm mode the elapsed time to complete an image decreased on average from 106 to 74 seconds compared to annotation mode. Simultaneously, the mean average precision (mAP) with an IoU > .5 increased from $\mu$=0.47 (min=0.29, max=0.68, $\sigma$=0.09) to $\mu$=0.55 (min=0.47, max=0.78, $\sigma$=0.8) and the average precision without grade from $\mu$=0.78 (min=0.47, max=0.91, $\sigma$=0.10) to $\mu$=0.92 (min=0.78, max=0.96, $\sigma$=0.04) (Fig ~\ref{0000-fig2}). Additionally, as shown in (Fig ~\ref{0000-fig2}) there was no obvious performance difference between the groups. If the participants performed the crowd-algorithm task prior to the annotation task, the accuracy variance decreased from 8\,\% (min=53, max=86, $\mu$=68) to 6\,\% (min=59, max=80, $\mu$=67), and the mAP variance decreased from 11\% (min=0.29, max=0.68, $\mu$=0.46) to 4\,\% (min=0.44, max=0.59, $\mu$=0.48). The participants found and corrected 74\,\% of the artificially removed cells and 86\,\% of the non-maximum suppression artefacts, but they failed to correct 67\,\% of the cells with artificially increased grade. 

The intra-observer cell-based classification accuracy ranged from 56-86\,\% ($\mu$=68, $\sigma$=8) with a mean Cohen's kappa score of 0.59. For the object detection performance the mAP ranges from 0.68-0.74 ($\mu$=0.46, $\sigma$=0.13) (Fig ~\ref{0000-fig1}).

The code for all experiments is available online (https://github.com/ ChristianMarzahl/EIPH\_WSI), together with the anonymised participant contributions and the image dataset.

\section{Discussion and outlook}

Our study shows that the use of precomputed annotations may lead to an increase in accuracy independent of the contributor's skill level, and to a reduction of interaction time by 30\%. Remarkably, only the two experts with the highest overall scores (a veterinary pathologist and a cytologist) deteriorated by around two percent in crowd-algorithm mode due to the manifested effect of accepting augmented grades. Although contributors were grouped and selected in concordance to their level of expertise with BAL cytology, there was no apparent difference between the performance of the groups. Additionally, we noticed a training effect when the crowd-algorithm mode was performed first, which was recognisable by a reduced variance and an increased mean average precision in the following annotation mode.

Participants achieved excellent results in the task of correcting non-maximum suppression artefacts or missing cells but failed to correct two thirds of the artificially increased cells. A reason for this could be that the Labelbox user interface adds a shading overlay over each cell altering its visual appearance, which led to an increase in interaction time because contributors often activated and deactivated the shading to better visualize the cell underneath. Furthermore, the task of assigning a grade to cells seems to be more challenging than only identifying them, as shown by the high mAP scores irrespective of the grade. The observed effect towards accepting the augmented grades was independent of the contributor's skill level and should be closely monitored to avoid introducing any unwanted bias into the dataset.

One limitation of this study is that the field of view contained only a limited number of cells, which is not comparable to the usual process of annotating 300 cells freely on a whole slide image. In this case, human performance is expected to decrease. Another drawback of the crowd-algorithm approach is that training data annotated from a pathologist has to be available in a sufficient quantity to train a deep learning-based method. 

The insights from this study allow us to effectively use crowd-algorithm collaboration in further work to label large high quality whole slide image datasets. However, the bias towards accepting precomputed grades has to be considered, and the training period of participants has to be extended. In conclusion, we would recommend using crowd-algorithm collaboration for the task of grading pulmonary hemosiderophages only if high quality precomputed annotations are available, and only for the task of correcting object detection and not for finding classification errors.  

\section{Acknowledgments}

\ack{We thank Labelbox for providing us with a education license, and we thank all contributors for making this work possible. CAB gratefully acknowledges financial support received from the Dres. Jutta \& Georg Bruns-Stiftung f\"ur innovative Veterin\"armedizin.}

\bibliographystyle{bvm2020}

\bibliography{0000}
\marginpar{\color{white}E\articlenumber} 
\end{document}